\documentclass[12pt]{iopart}
\usepackage{iopams}
\usepackage{graphicx}
\usepackage{dcolumn}
\usepackage{bm}
\usepackage{epsfig}
\begin{document}

\title[Two-Body Spectra of Harmonically Trapped Atoms]{Universal Two-Body Spectra of Ultracold Harmonically Trapped Atoms in Two and Three Dimensions}

\author{N T Zinner}
\ead{zinner@phys.au.dk}
\address{Department of Physics and Astronomy, Aarhus University, 
DK-8000 Aarhus C, Denmark}

\date{\today}

\begin{abstract}
We consider the spectrum of two ultracold harmonically trapped atoms interacting via short-range interactions. 
The Green's function approach is used to unify the two and three dimensional cases. We derive criteria for 
the universality of the spectrum, i.e. its independence of the details of the short-range interaction. 
The results in three dimensions are examplified for narrow $s$-wave Feshbach resonances and we show how
effective range corrections can modify the rearrangement of the level structure. However, this 
requires extremely narrow resonances or very tight traps that are not currently experimentally
available. In the two-dimensional case we discuss the $p$-wave channel in detail and demonstrate how
the non-universality of the spectrum arises within the Green's function approach. We then show that 
the spectrum is not particularly sensitive to the short-distance details in the case when the two-body interaction 
has a bound state.
\end{abstract}
\maketitle

\section{Introduction}
In an age of rapidly increasing computational power, exact methods and benchmark solutions continue
to have tremendous importance as a means of gauging numerical calculations and provide invaluable
analytical insights \cite{sutherland04}. Ultracold atomic gases have emerged as 
a field with great potential as a laboratory benchmark for many fields of physics. The extreme
control exercised over the systems in terms of trapping geometry and inter-atomic interactions
allow experimenters to prepare samples that simulate the intricacies of many different models
that are applied in other fields of physics and the term 'quantum simulator' is 
often applied \cite{bloch2008}. The clean conditions
provide hope of understanding some of the paradigmatic models such as the Hubbard model 
that are used frequently in the study of materials and other condensed-matter systems.

In the case of cold atoms, the interactions are typically short-ranged and the samples have to be 
in an external confinement, most often provided by a magnetic or optical potential. In the 
case of a harmonic oscillator trapping potential, it turns out that the problem of two
atoms interacting through a short-range potential can be exactly solved as demonstrated by
Busch {\it et al.} \cite{busch98}. The authors of Ref.~\cite{busch98} approximated the 
two-body interaction potential by a zero-range pseudopotential introduced long ago by 
Hellmann \cite{hellmann35} and Fermi \cite{fermi36}. The predictions of this exact model was subsequently tested experimentally
in an optical lattice and found to be a very accurate description of the two-atom system \cite{stoferle06}.

The pseudopotential approach has become somewhat of a paradigm itself in cold atoms. Its 
success can largely be attributed to a seperation of scales; 
the two-body collisions energy is small at the low temperatures one usually aims for, and 
the density of the system is also much lower than typical matter densities.
One can then model the interaction using only a few low-energy parameters such 
as the scattering length, $a$, and effective range, $r_e$ \cite{bethe49}. The true range of the 
potential, given by the van der Waals length for neutral atoms, is
much smaller than the interparticle spacing. The parameters $a$ and $r_e$ therefore
characterize the system, and since these are independent of the shape of the 
two-body potential, one refers to this as a universal regime, i.e. when $|a|,n^{-1/3}\gg r_0$
with $n$ the density. A really nice feature of the cold atomic gas system is the tunability 
of the interaction parameters through Feshbach resonances \cite{chin10}, which allows one to 
explore the full region of parameter space, including the interesting unitarity limit where
$1/|a|\to 0$.

In the case of a trapped system with harmonic oscillator length, $b$, a fundamental question 
concerns the binding energy of an $N$-body system in the universal regime when we also 
require that $r_0\ll b$. For equal mass particles, this has been studied numerically using 
a host of different methods \cite{dubois01}.
Interestingly, for $a\to \infty$, the three-body problem in a trap can be exactly solved \cite{werner06a}.
In the same spirit, exactly solvable models in a
harmonic approximation approach have been proposed \cite{magda00}. Very recently,
experiments in so-called microtraps have demonstrated that few-body systems can in fact be 
produced with cold atoms and universality and shell structure can be explored \cite{serwane11}.
The general framework of effective field theory is very suitable for problems
with seperation of scales, and it has been applied successfully to the
three-body problem in cold atoms \cite{braaten06}. Recently, there has been a lot of interest
in applying these techniques within an oscillator basis \cite{haxton02,stetcu07a}
to address few-body Fermi systems in cold atoms and in nuclei within the no-core shell-model
approach \cite{stetcu07b}. Similar methods have also been used to study few-body bosonic 
systems in traps (see \cite{tolle2011} and \cite{blume2012} for details and references).

The model of Busch {\it et al.}
is the basic foundation upon which many of the developments discussed above reside.
Here we take a fresh look at the model from a Green's function 
point of view \cite{castin07,idzi06,yip08}. This is done within a two-channel formalism.
We consider the general angular momentum $l$-wave case and 
derive the expression for the eigenspectrum in three dimensions as a function of the scattering phase shift. 
For $l\geq 2$, there are obstructing terms that imply a dependence on the short-range two-body potential, i.e.
the spectrum is non-universal. A criterion for the applicability of the universal formula is subsequently 
derived. As an example, we consider a two-channel model for narrow Feshbach resonances and show that interesting
spectral changes occur when including the effective range term. 
This can be interpreted in terms of the Zeldovich 
rearrangement effect \cite{zeldo60} which occurs in systems with a long-range attractive potential 
and a short-range two-body attraction that dominates at small distance (some recent discussion of the
effect can be found in Refs.~\cite{richard07,farrell11}).
The observation of these effects in experiments require, however, extremely tight 
trapping potentials or extremely narrow Feshbach resonances, both of which are beyond the current experimental
capabilities.

Two-dimensional setups are currently of great interest in 
the cold atom community and a number of impressive experimental results have been reported recently
\cite{kuhnle10}.
In the second part of the paper we consider a two-dimensional geometry and derive the eigenspectrum 
within the Green's function approach.  
We find that for angular momentum $m\geq 1$, there are 
non-universal terms, i.e. a spectrum that depends on the short-range details of the two-body potential.
Emphasis is put on the $p$-wave case $m=1$, where we find a closed formula for the spectrum in terms
of the low-energy parameters of the interaction. To obtain the eigenspectrum, we use a generic
form of the $p$-wave phase-shift in two dimensions, which is similar to that obtained in hard-sphere
or square well potential models. The $p$-wave spectrum in the case where the interaction allows a 
two-body bound state is very similar to that of $s$-waves. The shape of the $p$-wave spectrum is
almost universal, depending only slightly on the exact potential 
model used.

\section{Basic Two-Channel Formalism}
The physics around Feshbach resonances \cite{chin10} is most naturally by models which explicitly
take the open (scattering) channel and the closed (molecular) channel into account \cite{kohler06,chin10}.
Here we consider such a model within the Green's function approach \cite{castin07}.
The setup has a  dressed state with open, $\Psi(\bm r)$, and closed channel, $\beta\phi(\bm r)$ wavefunctions, where $\beta$ is the amplitude of the closed channel and the normalization is $|\beta|^2+\int d{\bm r}|\Psi(\bm r)|^2=1$ since we assume that $\phi$ is normalized. The Hilbert space of the closed channel is therefore one-dimensional for simplicity. Notice that $\phi(\bm r)$ has angular momentum $l$ and projection $m_l$.
The wave eqautions are 
\begin{eqnarray}
&\mathcal{D}_{trap}\Psi(\bm r)+W(\bm r)\beta\phi(\bm r)=E\Psi(\bm r)&\label{eq1}\\
&W(\bm r)\Psi(\bm r)+E_{mol}\beta\phi(\bm r)=E\beta\phi(\bm r),&\label{eq2}
\end{eqnarray}
where the operator is $\mathcal{D}_{trap}=\mathcal{D}_{free}+V(\bm r)$. Here $V(\bm r)$ is the trapping potential which we assume to be an isotropic harmonic oscillator with trap length $b=\sqrt{\hbar/\mu\omega}$ where $\omega$ is the oscillator frequency and $\mu$ is the reduced mass, i.e. $V(\bm r)=\frac{1}{2}\mu\omega^2\bm r^2$ The free particle operator has the standard form $\mathcal{D}_{free}=-\hbar^2\vec{\nabla}^2/2\mu$. The energy of the closed channel molecule is denoted $E_{mol}$ and the coupling between the channels is parametrized by the real function $W(\bm r)$. We assume
that the range of $W(\bm r)$ is much smaller than $b$.

\section{Three Dimensions}
The 3D Green's function $G_{E}(\bm r,\bm r')$ is defined by
\begin{eqnarray}
\left[\mathcal{D}_{trap}-E\right]G_{E}(\bm r,\bm r')=\frac{2\pi\hbar^2}{\mu}\delta(\bm r-\bm r').
\label{green}
\end{eqnarray}
If we define
\begin{eqnarray}
\mathcal{F}=E-E_{mol}+\frac{\mu}{2\pi\hbar^2}\int d{\bm r}d{\bm r'} \tilde\phi^*(\bm r)G_{E}(\bm r,\bm r')\tilde\phi(\bm r'),
\end{eqnarray}
where $\tilde\phi(\bm r)=W(\bm r)\phi(\bm r)$, then upon
substitution of Eq.~\ref{green} into Eqs.~\ref{eq1} and \ref{eq2} we obtain $\mathcal{F}=0$.
In momentum space we have
$\tilde \phi(\bm k)=\sqrt{4\pi}i^l k^l Y_{lm}(\hat{\bm k}) \alpha_{lm}$, which defines the coupling constant \cite{yip08}
\begin{eqnarray}\label{alphadef}
\alpha_{lm}=\sqrt{4\pi}k^{-l}\int d{\bm r} \tilde\phi(\bm r)j_l(kr)Y_{lm}^{*}({\bm r}).
\end{eqnarray}
In order to relate the spectrum in the trap to the free-particle
scattering properties we have to calculate the scattering amplitude. Therefore we consider the scattering problem
\begin{eqnarray}
\Psi(\bm r)=e^{i\bm k \cdot \bm r}-\frac{\mu}{2\pi\hbar^2}\beta\int d{\bm r'}\tilde\phi(\bm r')G_{E}^{0}(\bm r,\bm r'),\label{scat}
\end{eqnarray}
where $G_{E}^{0}(\bm r,\bm r')$ is the free-particle Green's function and $E=\hbar^2 k^2/2\mu$ is the scattering energy (see \ref{appg}).
Solving for $\beta$ in Eq.~\ref{eq2} and inserting into the scattering solution in Eq.~\ref{scat}, we find
\begin{eqnarray}
\Psi(\bm r)=e^{i\bm k \cdot \bm r}-\frac{\mu}{2\pi\hbar^2}\frac{\tilde\phi^*(\bm k)\int d{\bm r'}\tilde\phi(\bm r')G_{E}^{0}(\bm r,\bm r')}{\mathcal{F}_0},
\end{eqnarray}
where $\mathcal{F}_0$ is defined analogous to $\mathcal{F}$ above but with the free Green's function $G^{0}_{E}(\bm r,\bm r')$.
The solution of Eq.~\ref{green} for out-going wave boundary conditions when the trap is absent has the asymptotic behavior
\begin{eqnarray}
G_{E}^{0}(\bm r,\bm r')=\frac{e^{ik\vert \bm r-\bm r'\vert}}{\vert \bm r-\bm r'\vert}\to \frac{e^{ikr}}{r}e^{-i\bm k'\cdot \bm r'}\, \textrm{for}\,r\to \infty,
\end{eqnarray}
where $\bm k'$ is the final momentum which fulfills $\vert \bm k\vert=\vert \bm k'\vert$. Using the fact that 
\begin{eqnarray}
\tilde\phi^*(\bm k)=\int d\bm r e^{-i\bm k\cdot \bm r}\tilde\phi^*(\bm r),
\end{eqnarray}
we finally obtain
\begin{eqnarray}
f_{\bm k}(\bm k')=\frac{-\frac{\mu}{2\pi\hbar^2}\tilde\phi^*(\bm k)\tilde\phi(\bm k')}
{\mathcal{F}_0}.
\end{eqnarray}
Through partial wave decomposition 
$f_{\bm k}(\bm k')=4\pi \sum_{lm}f_{lm}(k) Y_{lm}^{*}({\bm k}) Y_{lm}({\bm k'})$ we get
\begin{eqnarray}
\frac{|\alpha_{lm}|^2k^{2l}}{f_{lm}(k)}=
&\frac{2\pi\hbar^2}{\mu}\left(E_{mol}-E\right)&\nonumber\\&-\int d{\bm r}d{\bm r'} \tilde\phi^*(\bm r)G_{E}^{0}(\bm r,\bm r')\tilde\phi(\bm r').&
\end{eqnarray}

Since both $G_{E}^{0}(\bm r,\bm r')$ and $G_{E}^{}(\bm r,\bm r')$ are singular at $\bm r=\bm r'$, we have to regularize by isolating the finite part
through $G_{E}^{R}(\bm r,\bm r')=G_{E}^{}(\bm r,\bm r')-G_{E}^{0}(\bm r,\bm r')$. We find
\begin{eqnarray}\label{ceq}
\frac{|\alpha_{lm}|^2 k^{2l}}{f_{lm}(k)}=\int d{\bm r}d{\bm r'} \tilde\phi^*(\bm r)G_{E}^{R}(\bm r,\bm r')\tilde\phi(\bm r'),
\end{eqnarray}
where $G_{E}^{R}(\bm r,\bm r')$ satisfies
\begin{eqnarray}\label{inhom}
\left(\mathcal{D}_{trap}-E\right)G_{E}^{R}(\bm r,\bm r')=-V(\bm r)G_{E}^{0}(\bm r,\bm r').
\end{eqnarray}
By symmetry we only need to consider $r>r'$, and we therefore write 
\begin{eqnarray}
G_{E}^{R}(\bm r,\bm r')=g_l(r,r')j_l(kr')Y_{lm}^{*}(\hat{\bm r})Y_{lm}^{}(\hat{\bm r'}),\,\, r>r'.
\end{eqnarray}
The solution for $g_l(r,r')$ can easily obtained by noting that a particular solution to 
Eq.~\ref{inhom} is $-G_{E}^{0}(\bm r,\bm r')$ \cite{yip08}. Adding the homogenoues solution gives
\begin{eqnarray}\label{gsol}
&g_l(r,r')=e^{-\frac{r^2}{2b^2}}\left[A(r')(\frac{r}{b})^l M(-\mu_1,l+3/2;r^2/b^2)\right.&\nonumber\\
&\left.+B(r')(\frac{b}{r})^{l+1}M(-\mu_2,1/2-l;r^2/b^2)\right]-4\pi i k h_{l}^{(1)}(kr),&
\end{eqnarray}
with $\mu_1=\frac{E}{2\hbar\omega}-l/2-3/4$ and $\mu_2=\frac{E}{2\hbar\omega}+l/2-1/4$. 
Here $M(a,b;x)$ is the confluent hypergeometric function and $h^{(1)}_{l}(x)$ is the 
spherical Hankel function of the first kind.
Demanding that $G_{E}^{R}(\bm r,\bm r')$ vanish as $r\rightarrow\infty$ yields 
\begin{eqnarray}\label{aform}
A(r')=-\frac{\Gamma(1/2-l)\Gamma(-\mu_1)}{\Gamma(l+3/2)\Gamma(-\mu_2)}B(r').
\end{eqnarray}
Furthermore, by demanding that $G_{E}^{R}(\bm r,\bm r')$ be regular as $r,r'\rightarrow 0$ (while maintaining
the condition $r>r'$), we demand that
\begin{eqnarray}\label{bform}
\lim_{r'\rightarrow 0}B(r')=\frac{\Gamma(l+1/2)}{b}\frac{2^{l+2}\sqrt{\pi}}{(kb)^l}.
\end{eqnarray} 
$g_l(r,r')$ can now be determined for small $r$.
For $r\rightarrow 0$ and $l\leq 2$
\begin{eqnarray}\label{gapp}
g(r,r')\approx\frac{4\pi (kr)^l}{(2l+1)!!k^{2l}} \left[\frac{(2l+1)!!A(r')k^l}{4\pi b^l}-ik^{2l+1} \right],
\end{eqnarray}
where the leading term $j_l(kr)$ for $kr\ll 1$ has been isolated to resemble the structure of 
Eq.~\ref{alphadef}.
The imaginary part of Eq.~\ref{gapp} comes directly from $\textrm{Im}\left[h^{(1)}_{l}(kr)\right]$.
For $l>2$ there are additional terms at order $r^a$ with $a<l$.
Inserting the solution into Eq.~\ref{ceq} and assuming that
$\tilde\phi$ is short-ranged, we find
\begin{eqnarray}\label{fres}
\frac{f_{lm}(k)}{k^{2l}}=\left[\frac{(2l+1)!!A(0)k^l}{4\pi b^l}-ik^{2l+1} \right]^{-1},
\end{eqnarray}
where $\alpha_{lm}$ cancels which was the object of the regularization procedure. Notice that this 
does not depend on $m$ due to the isotropy of the trap potential. 
Using Eqs.~\ref{aform} and \ref{bform}, we finally have
\begin{eqnarray}\label{formula}
\frac{f_{lm}(k)}{k^{2l}}=\left[\frac{(-1)^{l+1}2^{2l+1}}{b^{2l+1}}\frac{\Gamma(-\mu_1)}{\Gamma(-\mu_2)}-ik^{2l+1} \right]^{-1}.
\end{eqnarray}
This has to be related to the scattering amplitude in terms of the $l$-wave phase-shift, $\delta_l(k)$, which is 
\begin{eqnarray}
f_{lm}(k)=\frac{k^{2l}}{k^{2l+1}\cot\delta(k)-ik^{2l+1}}.
\end{eqnarray}
Matching with the solution above we obtain
\begin{eqnarray}\label{standard}
\frac{\Gamma(\frac{3}{4}+\frac{l}{2}-\frac{E}{2\hbar\omega})}{\Gamma(\frac{1}{4}-\frac{l}{2}-\frac{E}{2\hbar\omega})}=\frac{(-1)^{l+1}}{2^{2l+1}}(kb)^{2l+1}\cot\delta(k),
\end{eqnarray}
which recovers previous results \cite{busch98,yip08,suzuki09}.
Furthermore, the expression in Eq.~\ref{standard} shows how to include higher order terms from effective-range expansions. 

The result in Eq.~\ref{fres} holds for short-ranged $\tilde\phi$ and for $l<2$.
There are generally $r^a$ terms obstructing the simple formula derived above for $l\geq2$. These have order
from $a=3-l$ to $a=l-1$ in steps of two up to the leading $A(0)$ term of order $l$. For instance the $l=2$ case has a term
proportional to $r$, the $l=3$ case has a constant and an $r^2$ term and so forth. This can be seen by considering the 
series expansion of $h^{(1)}_{l}(x)$.
Since we demand that $\tilde\phi$
is very short-ranged, we assume $r/b\ll 1$. In general the most important term
near $r=0$ will be $r^{3-l}$ for $l\geq2$, thus it diverges for $l>3$ which must be compensated by the behavior of $\tilde\phi$.
A sensible criterion for this to happen is that $A(0)$ should 
dominate over the most divergent obstructing term. For general $l\geq 2$, this gives us the inequality
\begin{eqnarray}
|A(0)\frac{r^l}{b^l}| \gg |\frac{2^{l}\sqrt{\pi}\Gamma(l+1/2)}{(2l-3)(kb)^l}\left(\frac{r}{b}\right)^{3-l}|
\end{eqnarray}
$W(r)$ is assumed to be of short-range and $\phi(r)$ is an $l$-wave wave function. 
Thus there will be some intermediate region
around $r=r_0$ where $\tilde \phi$ has its weight. We want the condition to be satisfied at this distance.
Inserting 
$A(0)$ and using Eq.~\ref{standard} we arrive at
\begin{eqnarray}\label{cond}
\left(\frac{r_0}{b}\right)^{2l-3}\gg \frac{2l+1}{2l-3}\frac{[(2l-1)!!]^2}{4}\frac{|\tan\delta_l(k)|}{(kb)^{2l+1}},
\end{eqnarray}
which agrees with Ref.~\cite{suzuki09} for $l=2$. The distance $r_0$ is essentially the same as the matching distance of Ref.~\cite{jonsell02} and Ref.~\cite{suzuki09}. We still require $r_0/b\ll 1$, so when the right-hand side blows up, the universal formula breaks down and 
the details of the two-body interaction become important.

For concreteness, let us consider a model for the coupling between the channels where $W(r)=W_0e^{-r/a_0}$ with $a_0$ the Bohr radius \cite{nygaard06}. Assuming that $\phi(r)\propto r^l$, the maximum of $\tilde\phi$ is at $r_0=la_0$. The left-hand side of Eq.~\ref{cond} is therefore very small as $b\sim 1\mu$m for typical traps. The factors depending on $l$ on the right-hand side are increasing but only gradually. 
Using the lowest-order in the effective-range expansion, the criterion is $(a_l/b)^{2l+1}\ll 1$ where $a_l$ is the $l$-wave scattering length (whenever it is well-defined). Thus we conclude that the formula works only away from resonance.

\begin{figure}[t!]
\centering
  \epsfig{file=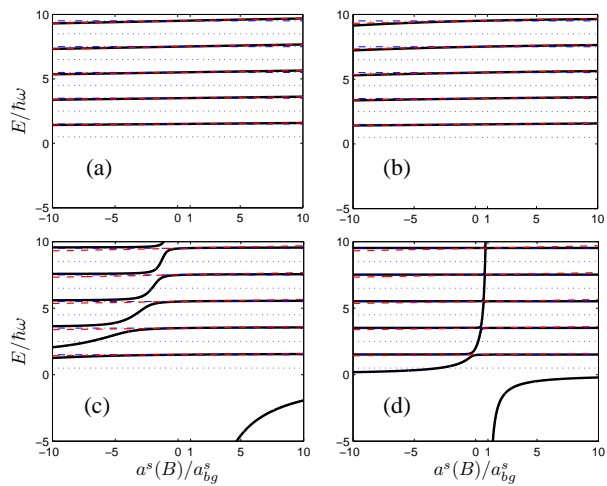,clip=true,scale=0.45}
  \caption{(Color online) 3D $s$-wave spectrum with $a_{bg}^{s}/b=0.01$ and (a) $r_{e0}^{s}/b=0.1$, (b) $r_{e0}^{s}/b=1$, (c) $r_{e0}^{s}/b=10$, and (d) $r_{e0}^{s}/b=100$. The solid (black) lines are the full solutions while the dashed (red) lines are for $r_{e0}^{s}=0$. The dot-dashed (blue) horizontal lines indicate the non-interacting level structure (visible in the top left corner of (a) and (b)), while the dotted (blue) vertical lines are asymptotes $1\hbar\omega$ above (or below) the non-interacting levels.}
  \label{fig1}
\end{figure}

\subsection{3D Feshbach Model}
The formula in Eq.~\ref{standard} has been discussed by a number of authors for both 
$s$-waves \cite{busch98,blume2002,jonsell02}, $p$-waves \cite{idzi06,yip08}, and $d$-waves \cite{suzuki09}. 
In the context of Feshbach resonances, most studies have assumed that the effective range can be neglected which
is true for wide resoanances \cite{chin10}. To complement this, we considered here the opposite
limit of very narrow resonances where the effective-range is large. This is interesting in cold 
atomic gases with two or more different species of atoms which typically have narrow resonances. In particular,
at the point where the scattering length goes to zero we expect the corrections from higher-order terms to 
become important \cite{zinner2009a}.
The universal behavior of one-channel models is described through one parameter, the scattering length. As a function of applied 
external magnetic field we parametrize the $l$-wave field dependnent scattering length $a_l(B)$ in the following way
\begin{eqnarray}\label{fscat}
a^l(B)^{2l+1}=(a_{bg}^{l})^{2l+1}\left[ 1-\frac{\Delta B}{B-B_0} \right],
\end{eqnarray}
where $a_{bg}^{l}$ is the background scattering length away from the resonance located at $B=B_0$ with width $\Delta B$.
Since we are interested in going beyond the one parameter description, we use a two-channel model of a Feshbach resonance \cite{bruun05}
with corresponding open-open channel $s$-wave $T$-matrix
\begin{eqnarray}
T_{oo}^{s}=\frac{\frac{2\pi\hbar^2a_{bg}^{s}}{\mu}}{\left(1+\frac{\Delta\mu\Delta B}{\frac{\hbar^2k^2}{2\mu}-\Delta\mu(B-B_0)}\right)^{-1}+ia_{bg}^{s}k},
\end{eqnarray}
where $\Delta\mu$ is the difference in magnetic moment of the open and closed channels. 
Combining this with Eq.~\ref{formula}, we get
\begin{eqnarray}
\frac{b}{a_{bg}^{s}}(1+\frac{\Delta\mu\Delta B}{E-\Delta\mu(B-B_0)})^{-1}
=\sqrt{2}\frac{\Gamma(\frac{3}{4}-\frac{E}{2\hbar\omega})}{\Gamma(\frac{1}{4}-\frac{E}{2\hbar\omega})},
\end{eqnarray}
Introducing the background effective range $r_{e0}^{s}=-\hbar^2/(\Delta\mu\Delta B \mu a_{bg}^{s})$, and the
useful quantities $x=(B-B_0)/\Delta B$ and $f(E)=\sqrt{2}\Gamma(\frac{3}{4}-\frac{E}{2\hbar\omega})/\Gamma(\frac{1}{4}-\frac{E}{2\hbar\omega})$, we have 
\begin{eqnarray}
\left(1+\frac{1}{\frac{2b^2}{a_{bg}^{s}|r_{e0}^{s}|}\frac{E}{\hbar\omega}-x}\right)^{-1}=\frac{a_{bg}^{s}}{b}f(E).
\end{eqnarray}
We note that since $a_{bg}^{l}\Delta\mu\Delta B>0$ for all resonances with any $l$ \cite{chin10}, we have $r_{e0}^{l}<0$ always. Isolating $x$
yields 
\begin{eqnarray}\label{eigs}
x=\frac{1}{2}\frac{a_{bg}^{s}}{b}\frac{|r_{e0}^{s}|}{b}\frac{E}{\hbar\omega}-\frac{\frac{a_{bg}^{s}}{b}f(E)}{1-\frac{a_{bg}^{s}}{b}f(E)}.
\end{eqnarray}
In the case of a wide resonance or a large trap, $|r_{e0}^{s}|/b \rightarrow 0$, we recover $\frac{b}{a^{s}(B)}=f(E)$.

There is one caveat that has to be addressed before we proceed to study the effective range corrections to the 
two-body spectrum. This is related to the limit when the scattering length becomes where small. Here it is
not necessarily clear that the properties of the spectrum will be universal in the sense that higher-order 
parameters from the effective range expansion (beyond $a$ and $r_e$) can be neglected. This issue has been 
discussed in the context of effective field theory in Ref.~\cite{kolck1999}. 

The details of Feshbach resonances
when the scattering length goes to zero has been considered for both trapped bosons and fermions
in Ref.~\cite{zinner2009a}. There it was found that the effective interaction is quadratic in 
the relative momentum at lowest order, since the usual constant piece proportional to $a^{s}$
vanishes. The coefficient depends on the background parameters of the resonance through the 
combination $(a_{bg}^{s})^2r_{e0}$. The studies in Ref.~\cite{zinner2009a} demonstrated that 
no anomalous behavior is seen when approaching the zero-crossing of the Feshbach resonance.

Of course, if this quantity happens to be very small, even
higher-order terms in the effective-range expansion must be taken into account. Here we 
are assuming that $|a_{bg}^{s}|\gg r_\textrm{vdW}$ is much larger than the true range of the potential, given 
by the van der Waals length, $r_\textrm{vdW}$. Likewise, for the narrow resonances we are interested
in here $|r_{e0}|\gg r_\textrm{vdW}$. It is in this regime that we expect the behavior to be 
universal, since this implies that low-energy scattering still dominates the two-body collisional
dynamics. The smallest values used below are $a_{bg}^{s}/b=0.01$. For a typical trap with $b\sim 1\mu$m, 
this is larger than $r_\textrm{vdW}$ for most atoms used in cold gas experiments.

\begin{figure}[t!]
\centering
  \epsfig{file=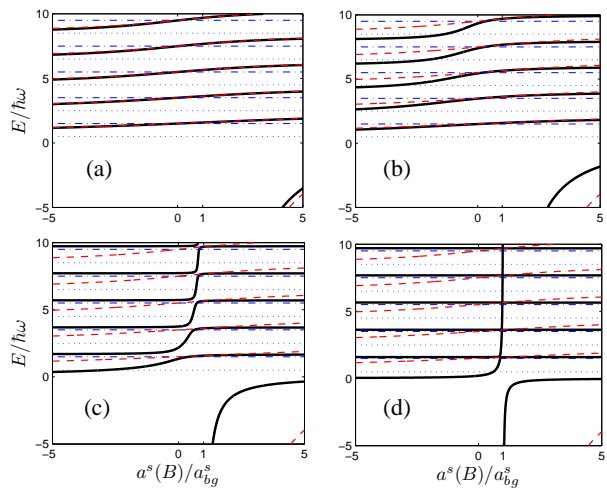,clip=true,scale=0.45}
  \caption{(Color online) Same as Fig.~\ref{fig1} for $a_{bg}^{s}/b=0.1$ and (a) $r_{e0}^{s}/b=0.1$, (b) $r_{e0}^{s}/b=1$, (c) $r_{e0}^{s}/b=10$, and (d) $r_{e0}^{s}/b=100$.}
  \label{fig2}
\end{figure}

\begin{figure}[t!]
\centering
  \epsfig{file=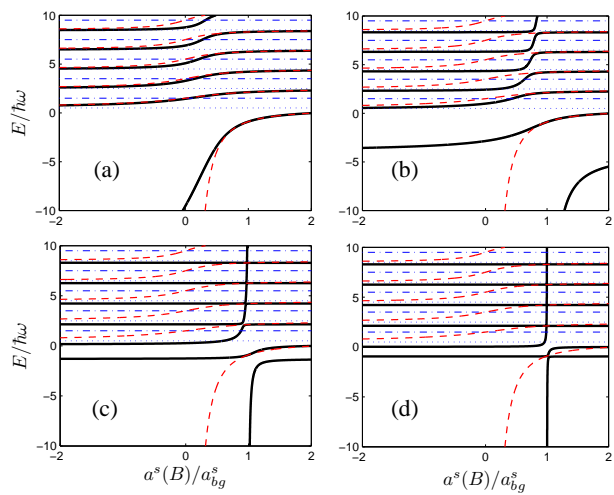,clip=true,scale=0.45}
  \caption{(Color online) Same as Fig.~\ref{fig1} for $a_{bg}^{s}/b=1$ and (a) $r_{e0}^{s}/b=0.1$, (b) $r_{e0}^{s}/b=1$, (c) $r_{e0}^{s}/b=10$, and (d) $r_{e0}^{s}/b=100$.}
  \label{fig3}
\end{figure}

In Figs.~\ref{fig1}, \ref{fig2}, and \ref{fig3} the two-body spectrum in the trap is plotted for $s$-wave interactions and various values of the background parameters of the resonance, $a_{bg}^{s}$ and $r_{e0}^{s}$, as funciton of $a^s(B)/a_{bg}^{s}$. We plot both the full
solution and the standard case with $r_{e0}=0$ for comparison. In Fig.~\ref{fig1}, $a_{bg}^{s}/b=0.01$ which is very small. This means that the spectrum is almost equal to the non-interacting case when $r_{e0}^{s}$ is also small. Note that the molecular state has energy proportional to $-(a_{bg}^{s})^{-2}$ so its energy is below the range of the figure in the case of $r_{e0}^{s}=0$. In Fig.~\ref{fig1}(c) and (d) the situation changes and a molecular state can be seen. One also clearly see the Zeldovich rearrangement effect \cite{zeldo60,richard07} of the levels in the right of (c) and middle of (d). The connection between the Busch model and this effects in the $r_{e0}^{s}=0$ case was discussed recently by Farrell {\it et al.} \cite{farrell11}. However, the $a^s(B)$ value of the rearrangements can now depend on the level since we have the term linear in $E$ in Eq.~\ref{eigs}. In contrast, for $r_{e0}^{s}=0$ the rearrangments happens at $a^s(B)=0$.

In Figs.~\ref{fig2} and \ref{fig3} we exhibit the spectrum for larger values of $a_{bg}^{s}/b$, which means that the molecular state is now seen even for small $r_{e0}^{s}$. From these figures it is also clear that for large $r_{e0}^{s}/b$ the rearrangement happens when $a^s(B)=a_{bg}^{s}$. This can again be understood from Eq.~\ref{eigs} since $x\to \infty$ when $a^s(B)\to a_{bg}^{s}$ and which implies $1=a_{bg}^{s}f(E)/b$. The presence of the linear $E$ term for $r_{e0}^{s}>0$ gives distortion to this simple picture and enriches the rearrangement effect. Notice that in Fig.~\ref{fig3}(a) the lowest state shown is in fact not the molecular state but the first excited state. As $r_{e0}^{s}$ increases the molecular state is pulled up in energy as seen in Fig.~\ref{fig3}(b). Also, panel (d) demonstrates that for very large values of $r_{e0}^{s}$, the region where the the levels rearrange can become very small, yielding almost abrupt jumps in the spectrum.

To access the spectra above experimentally will require large values of $a_{bg}^{s}/b$ and $r_{e0}^{s}/b$. For typical traps with $b$ of order
$\mu$m, this seems inaccessible. However, a single site of an optical lattice could have a much smaller $b$ and has been used before to probe the two-body spectrum \cite{stoferle06}. Assuming that one could make large reduction in $b$, we still require Feshbach resonances with large background parameters. Atoms like $^{23}$Na or $^{87}$Rb do in fact have known Feshbach resonances that are extremely narrow \cite{chin10} and will give $r_{e0}$ of order $\mu$m. Resonances between two different mass atoms also tend to be narrow in general so mixtures is an option. However, narrow resonances require the ability to tune the magnetic field extremely precisely. The level of tunability required here is probably beyond any current experiment but will perhaps be available in next generation experimental setups.

\section{Two Dimensions}
The two-dimensional case is similar but contains the peculiarities of 2D scattering \cite{adhikari86}. The free 2D Green's function 
with out-going boundary condition is
\begin{eqnarray}
G_{E}^{0}(\bm r,\bm r')=-i\frac{\mu}{2\hbar^2}H_{0}^{(1)}(k\vert \bm r-\bm r'\vert),
\end{eqnarray}
where $H_{m}^{(1)}(x)$ is the $m$'th order Hankel function of the first kind. We use the 
partial wave decomposition of the Green's function (see \ref{appg})
\begin{eqnarray}
G_{E}^{0}(\bm r,\bm r')=i\pi\sum_{m=-\infty}^{\infty} J_{|m|}(kr')H_{|m|}^{(1)}(kr),
\end{eqnarray}
for $r>r'$. $J_{m}(x)$ is the Bessel function of order $m$.
The coupling constant, $\alpha_m$, has to be modified slightly to fit the 2D geometry. We define
\begin{eqnarray}
\alpha_m=\frac{\sqrt{\pi}}{k^{|m|}}\int d\bm r J_{|m|}(kr)e^{im\theta_{r}}\tilde\phi(\bm r),
\end{eqnarray}
where $\tilde\phi(\bm r)=\phi(\bm r)W(\bm r)$ just like in the 3D case above and $\theta_r$ is 
the angle of the 2D vector $\bm r$. For a wave function with angular momentum $m$, the 
Fourier transform becomes
\begin{eqnarray}
\tilde \phi(\bm k)=\frac{1}{\sqrt{\pi}}\sum_{m=-\infty}^{\infty} i^m k^{|m|}\alpha_m e^{im\theta_k}.
\end{eqnarray}
The finite part of the 2D Green's function has to fulfill
\begin{eqnarray}\label{2Dgreen}
\frac{|\alpha_{m}|^2 k^{2m}}{f_{m}(k)}=\int d{\bm r}d{\bm r'} \tilde\phi^*(\bm r)G_{E}^{R}(\bm r,\bm r')\tilde\phi(\bm r'),
\end{eqnarray}
where $G_{E}^{R}(\bm r,\bm r')$ satisfies once again Eq.~\ref{inhom} and the scattering amplitude, $f_m(k)$,
is connected to the scattering phase shift, $\delta_m(k)$, through \cite{adhikari86}
\begin{eqnarray}
f_m(k)=\frac{k^{2|m|}}{k^{2|m|}\cot\delta_m(k)-ik^{2|m|}}.
\end{eqnarray}
This can be solved similarly to the 3D case by assuming that ($r>r'$)
\begin{eqnarray}
G_{E}^{R}(\bm r,\bm r')=G_{|m|}(r,r')J_{|m|}(kr')e^{im\theta_r-im\theta_{r'}},
\end{eqnarray}
which yields 
\begin{eqnarray}
G_{|m|}(r,r')=&A(r')U[-\nu_m,|m|+1,r^2/b^2]\left(\frac{r}{b}\right)^{|m|}e^{-\frac{r^2}{2b^2}}&\\\nonumber
&-i\pi H_{|m|}^{(1)}(kr),&
\end{eqnarray}
where we define $\nu_m$ through $E=\hbar\omega(2\nu_m+|m|+1)$.
Here we have taken a small shortcut by introducting the Tricomi hypergeometric function, $U(a,b,z)$, which
is the convergent solution for $z\gg 1$. Demanding that $G_{E}^{R}(r,r')$ be regular at the origin yields
the condition
\begin{eqnarray}
\lim_{r'\to 0}A(r')=\frac{2^{|m|}\Gamma[-\nu_m]}{k^{|m|}}.
\end{eqnarray}
Proceeding with general $m$ is not attractive since the expressions for the lowest order terms are 
cumbersome. We therefore specialize to specific $m$ values.

\begin{figure}[t!]
\centering
  \epsfig{file=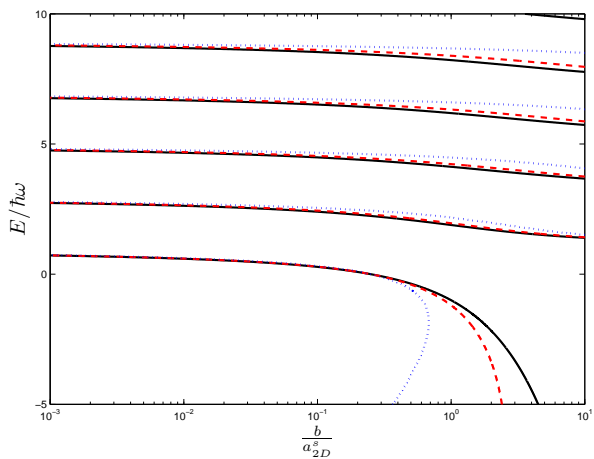,clip=true,scale=0.40}
  \caption{(Color online) Two-body spectrum in 2D for particles interacting in the $s$-wave channel. Solid (black) line is for $r_e/b=0$, dashed (red) for $r_e/b=0.5$, and dotted (blue) has $r_e/b=1.0$.}
  \label{fig4}
\end{figure}

For $m=0$, the spectrum should be universal according to the pseudopotential approach \cite{busch98,blume2006}. 
The Green's function approach has been discussed in a quasi-2D geometry with a tight transverse confinement \cite{petrov2001}.
First consider the behavior of $G_0(r,r')$ at the origin
\begin{eqnarray}
G_0(r,r')\to -i\pi + 2 \ln\left[\frac{kb}{2}\right]-\psi\left(\frac{1}{2}-\frac{k^2}{4}\right),
\end{eqnarray}
for $r,r'\to 0\,(r>r')$, where $\psi(x)$ is the digamma function. 
The scattering phase shift for $m=0$ in 2D can be written \cite{verhaar84}
\begin{eqnarray}
\cot\delta_0(k)=\frac{2}{\pi}\left(\gamma+\ln\left[\frac{ka_{2D}^{s}}{2}\right]\right)+\frac{1}{2\pi}(r_{e}^{s})^{2}k^2,
\end{eqnarray}
where $a_{2D}^{s}$ is the 2D scattering length and $r_{e}^{s}$ is the effective range. $\gamma$ is Euler's constant. 
We have included the effective range term to discuss its effects below.
Using Eq.~\ref{2Dgreen}, we arrive at the eigenvalue equation for the spectrum
\begin{eqnarray}
\gamma+\frac{1}{2}\psi\left(-\nu_0\right)+\frac{(r_{e}^{s})^{2}}{b^2}\left(\nu_0+\frac{1}{2}\right)=\ln\left[\frac{b}{a_{2D}^{s}}\right].
\end{eqnarray}
In the case where $r_{e}^{s}=0$, this agrees with earlier work \cite{busch98,petrov2001,blume2006,liu2010,farrell10}. In the 
limit $\frac{b}{a_{2D}^{s}}\to \infty$, the energy
approaches the universal expression $E=-\hbar^2e^{-2\gamma}/2\mu [a_{2D}^{s}]^2$ (for $r_{e}^{s}=0$)
and represents the two-body bound state energy in the absence of the trap \cite{jensen2004}. This 
is reasonable since the trap becomes irrelevant for large binding energy and small bound state size.
The spectrum is shown in Fig.~\ref{fig4} for different values of $r_{e}^{s}$. We can see that
effective range corrections will alter the energetics of the lowest state quite severely within this
model. This is very similar to what is found for $p$-waves with range corrections in 3D \cite{idzi06}.

The Green's
function method is particularly transparent when including higher-order correction terms in comparison 
to the pseudopotential approach \cite{busch98} or, equivalently, 
the Bethe-Peiels boundary condition \cite{kartavtsev2006,liu2010}. A mathematical formulation of pseudopotential 
approaches in any dimension and for any angular momentum 
was recently discussed by Stampfer and Wagner \cite{stampfer2010} which details
the intricate problems of even dimensions in comparison to odd dimensions. The expression for the 
pseudopotential beyond lowest order is, however, involved. 
The Green's function approach accomplishes these corrections in a simple manner.

Whereas the case of $m=0$ was universal in the sense that the dependence on the details of the 
potential (through $\tilde\phi(\bm r$) dropped out, there are a lot of indications that the 
case $|m|>0$ is not so simple and will depend on short-distance physics. This has been 
pointed out by a number of authors, and leads to the introduction of energy-dependent scattering 
lengths \cite{wodkiewicz1991,blume2002,idzi06,blume2006} (see 
\cite{valiente11} for an alternative approach to pseudopotentials that can be applied in a well-defined
manner in momentum space in both 2D and 3D). We now
address this issue within the Green's function approach for $m=1$.

\begin{figure}[t!]
\centering
  \epsfig{file=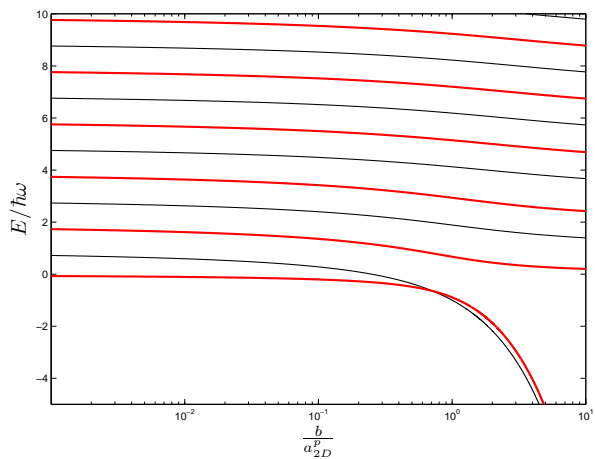,clip=true,scale=0.40}
  \caption{(Color online) Two-body spectrum in 2D for $p$-wave interactions. Dashed (red) line is $p$-wave with $r_{e}^{p}/b=0$. For comparison, the solid (black) line is $s$-wave with $r_{e}^{s}/b=0$}
  \label{fig5}
\end{figure}

The expansion of $G_1(r,r')$ is
\begin{eqnarray}
&G_1(r,r')\to \frac{\pi kr}{2}\left[ -i+\frac{2}{\pi}\ln\left[\frac{kb}{2}\right]+\frac{2}{\pi (kb)^2}\right.&\nonumber\\
&\left.-\frac{1}{\pi}\psi\left(-\nu_1\right)\right]&
\end{eqnarray} 
Here we have extracted a factor in front which agrees with $J_1(kr)$ to lowest order in $kr$. We find no other terms that
depend explicitly on $r$. 
In the $m=1$ channel it seems natural to define the phase-shift relation
that generalizes the $m=0$ result as (see \ref{appscat})
\begin{eqnarray}\label{genphase}
\cot\delta_1(k)=\frac{2}{\pi}\left(\gamma+\ln\left[\frac{ka_{2D}^{p}}{2}\right]\right)+\frac{A}{(ka_{2D}^{p})^2}+\frac{1}{2\pi}(r_{e}^{p})^2k^2,
\end{eqnarray}
to order $k^2$ with $A$ a dimensionless constant. The non-universal information about the 
two-body interaction potential is in fact carried by $A$ as pointed out in Ref.~\cite{randeria1990} and discussed further in
\ref{appscat}.
Let us consider the case when $A=0$ which occurs if there is a bound state at zero energy in 
the $m=1$ potential. 
In this case we arrive at
the very simple equation for the eigenspectrum
\begin{eqnarray}
\gamma+\frac{1}{2}\psi\left(-\nu_1\right)+\frac{r_{e}^{2}}{b^2}\left(\nu_1+1\right)+\frac{1}{4\nu_1+4}=\ln\left[\frac{b}{a_{2D}^{p}}\right].
\end{eqnarray}
The expression we get is very similar to the $m=0$ case, expect for the last term on the left-hand side which is 
the new piece. This extra term will vanish for large energies, but will be important around zero energy (where 
a bound state in free space resides). The spectrum is plotted in Fig.~\ref{fig5}. We can see that the levels are pushed
down compared to the $m=0$ case, and a state reside at zero energy in the non-interacting $\frac{b}{a_{2D}^{p}}\to 0$ limit. A 
bound state appearing below the $m=0$ ground state for $\frac{b}{a_{2D}^{p}}\to 0$ is not uncommon and occurs also for
$p$-waves in 3D \cite{idzi06}. In the case of $A<0$, there is a bound state with finite binding energy in the potential. 
However, we have checked that this only gives minor quantitative changes compared to the $A=0$ in Fig.~\ref{fig5}. Since
$A=0$ and $A<0$ yield qualitatively the same spectra, we conclude that the extra term proportional to $k^{-2}$ makes
little difference when the potential has a bound state (possibly at zero energy). Note that we are considering the 
zero-range limit for the potential, so there can be at most one bound state.

The $m=1$ spectrum in a 2D trap was discussed in Ref.~\cite{blume2006} and a spectrum can be found in Fig.~1 of Ref.~\cite{blume2006}.
We find good agreement with that result in the case when a bound state is present in the two-body potential, corresponding to 
$A<0$ in Eq.~\ref{genphase}. The level rearrangement takes place
at $a_{2D}^{p}\to 0$ as seen in Fig.~\ref{fig5} and there is a visible distortion of the manner in which this occurs similar
to the examples in 3D of Fig.~\ref{fig1}, \ref{fig2}, and \ref{fig3}. These features can also be seen in Ref.~\cite{blume2006}. 
When $A>0$ in Eq.~\ref{genphase}, we find that lowest state seen in Fig.~\ref{fig5} does not diverge to minus infinite 
in binding energy for $a_{2D}^{p}\to 0$, but rather behaves similar to the higher lying states. This is consistent with 
the finding in Ref.~\cite{blume2006}, although we caution that the scattering area used to parameterize the strength
in Ref.~\cite{blume2006} can have both signs while our $a_{2D}^{p}$ is defined to be positive.

\begin{figure}[t!]
\centering
  \epsfig{file=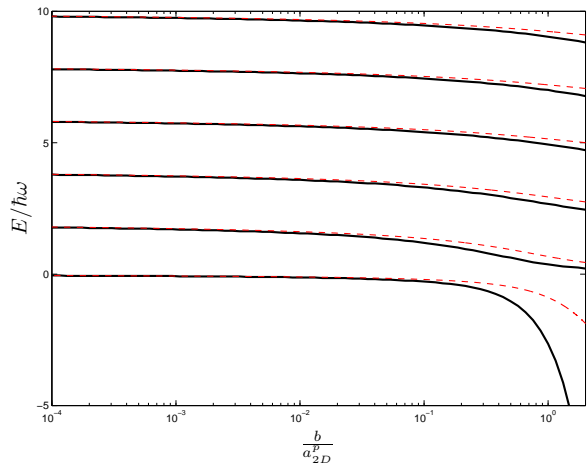,clip=true,scale=0.40}
  \caption{(Color online) 2D $p$-wave eigenspectrum using a phase-shift similar to the hard-sphere potential (solid (black) line). For 
  comparison, the spectrum using the generic phase-shift in Eq.~\ref{genphase} is shown by the dashed (red) line.}
  \label{fig6}
\end{figure}

To further explore the dependence of the $m=1$ spectrum on the potential parameters, we show in Fig.~\ref{fig6} results
for a phase-shift of the same form as the hard-sphere potential (\ref{appscat}) but with positive scattering 
length so that a bound state occurs. This is a somewhat unphysical potential but it helps illustrate the point that 
the spectrum is quite robust under changes in the value of $A$ (for $A\leq 0$) since Fig.~\ref{fig5} has $A=0$ 
while Fig.~\ref{fig6} has $A=-4/\pi$ (see Eq.~\ref{hardsphere}) . 
Using a square well instead yields almost identical 
results and we have not plotted this case. In fact, the procedure of using a model potential to fix the 
phase-shift used here is similar to the self-consistent energy-dependent pseudopotential methods employed
for $p$-waves in 3D in Ref.~\cite{idzi06}. This provides a significant improvement over the energy-independent
pseudopotential in that case.

In the case of $|m|>1$ we find that there are terms in $G_|m|(r,r')$ that do not vanish in the limit of $r,r'\to 0$
and that are not represented on the left-hand side of Eq.~\ref{2Dgreen}. We therefore conclude that the higher
partial waves in 2D also yield non-universal spectra.

\section{Conclusions and Outlook}
Using Green's function techniques we have derived the spectrum of two particles in an 
isotropic harmonic trap interacting through a potential that has a range that is much 
smaller than the trap length scale in both three and two dimensional space. In the 
3D case, the spectrum is universal, i.e. independent of the short-range details of 
the interaction, only for the $s$- and $p$-wave channels. For $d$-wave and beyond 
there is a dependence on the short-distance physics. We derive a general criterion
for when the universal expression is valid for higher angular momenta which implies
that one must be away from any resonances where the scattering length diverges.
In the 2D case, only the $s$-wave spectrum is truly universal, but we demonstrated
that the $p$-wave spectrum is not very sensitive to the exact details of the 
interaction as long as it can accomodate a two-body bound state.

The spectra in both 2D and 3D can be interpreted very nicely in terms of Zeldovich 
rearrangement, which occurs when adding a short-range (two-body interaciton) 
to a long-range potential (trap) and tuning through a resonance of the short-range part.
Including effective range corrections in the 3D $s$-wave channel shows that one 
can get a very rich set of rearrangment points depending on the background 
parameters of the resonance in the short-range interaction. In the realm of cold 
gases, this requires use of very narrow Feshbach resonances or very tight trapping
conditions, both of which are currently beyond experimental capabilities but 
hopefully could be explored in future generation experiments.

It would be interesting to extend the current formalism to polar molecules for
which external trapping potential are of course also always present in experiments. 
A number of recent works \cite{dipolarworks} have explored the bound state structure of such 
systems but it not clear how much influence a harmonic trap or an optical lattice
has on these few-body states.

\appendix

\section{Free Green's Functions}\label{appg}
The free Green's functions with out-going wave boundary condition 
used in this paper are taken to satisfy the equation 
\begin{eqnarray}\label{greendef}
\left[\frac{-\hbar^2\nabla^2}{2\mu}-E\right]G^{0}_{E}=\frac{2\pi\hbar^2}{\mu}\delta(\bm r-\bm r'),
\end{eqnarray}
where we define the corresponding momentum through $E=\hbar^2k^2/2\mu$.

\subsection{3D}
The solution in 3D is 
\begin{eqnarray}
G^{0}_{E}(\bm r,\bm r')=\frac{e^{ik|\bm r-\bm r|}}{|\bm r-\bm r'|}.
\end{eqnarray}
The partial wave expansion that we employ is written in terms of 
spherical Hankel functions of the first kind, $h_{l}^{(1)}(x)$, 
and spherical Bessel functions, $j_{l}(x)$. For $r>r'$ we have
\begin{eqnarray}
G^{0}_{E}(\bm r,\bm r')=4\pi i k \sum_{lm} j_{l}(kr')h_{l}^{(1)}(kr)Y_{lm}(\bm r)
Y_{lm}^{*}(\bm r'),
\end{eqnarray}
and for $r<r'$ one interchanges the radial variables.

\subsection{2D}
The solution in 2D has been given by many authors, here we follow Adhikari \cite{adhikari86}.
With out-going wave boundary condition, the solution is
\begin{eqnarray}
G^{0}_{E}(\bm r,\bm r')=-\frac{i\mu}{2\hbar^2}H_{0}^{(1)}(k|\bm r-\bm r'|), 
\end{eqnarray}
where $H_{m}^{(1)}(x)$ is the first Hankel function of order $m$. The partial
wave expansion of this Green's function appears to be less accessible and we 
therefore write it explicitly here. Starting from momentum space we define
$G_{E}^{0}=(E-H_0)^{-1}$. This implies that
\begin{eqnarray}
\langle \bm r |G_{0}^{E} | \bm r'\rangle =\frac{1}{(2\pi)^2}\int d^2 p
\frac{e^{i \bm p(\bm r -\bm r')}}{E-\frac{\hbar^2 p^2}{2\mu}},
\end{eqnarray}
where $p$ denotes the wave vector to be integrated over.
Using the expansion
\begin{eqnarray}
e^{ikx\cos\phi}=\sum_{m=-\infty}^{\infty}i^{|m|} J_{|m|}(kx)e^{im\phi},
\end{eqnarray}
where $J_m(x)$ is the Bessel function of order $m$,
we can turn this into
\begin{eqnarray}
-\frac{\mu}{\pi\hbar^2}\sum_{m} \int_{0}^{\infty} dp p \frac{J_{|m|}(pr)J_{|m|}(pr')}{p^2-k^2}.
\end{eqnarray}
This integral can be found in standard tables \cite{grad65} and through analytical contiuation 
we obtain for $r>r'$
\begin{eqnarray}
-\frac{i\mu}{2\hbar^2}\sum_m J_{|m|}(kr')H_{|m|}^{(1)}(kr).
\end{eqnarray}
With our normalization of the Green's function in Eq.~\ref{greendef}, we finally end up with
\begin{eqnarray}
G_{E}^{0}(\bm r,\bm r')=i\pi \sum_m J_{|m|}(kr')H_{|m|}^{(1)}(kr).
\end{eqnarray}

\section{2D Scattering}\label{appscat}
Scattering in 2D is complicated by the appearance of logarithmic terms in the typical 
wave function in the asymptotic region of large distance, 
which is a Neumann function, $Y_{m}(x)$. Note that 
we are only interesting in short-range potentials (vanishing for distances $r>r_0$)
for which the asymptotic solution is 
the free one. We can therefore write the angular momentum $m$ scattering wave 
function, $\Psi_m(r)$, for $r>r_0$ in the form
\begin{eqnarray}
\Psi_m(r)=A\left[\cot\delta_m(k)J_m(kr)-Y_m(kr)\right]
\end{eqnarray}
where $E=\hbar^2k^2/2\mu$ and $N$ is a normalization constant. The phase-shift $\delta_m(k)$ 
can be calculated from
\begin{eqnarray}
\cot\delta_m(k)=\frac{xY_{m}^{'}(x)-\gamma_m Y_{m}(x)}{xJ_{m}^{'}(x)-\gamma_m J_{m}(x)},
\end{eqnarray}
where $x=kr_0$ and prime denotes derivative with respect to $x$. The logarithmic derivative
is 
\begin{eqnarray}
\gamma_m=\left[\frac{1}{\Psi_m(r)}\frac{d\Psi_m(r)}{dr}\right]_{r=r_0}.
\end{eqnarray}

We would like to discuss this in terms of an appropiately defined scattering length. Here
we follow the intuitively clear defintion \cite{nielsen99}
\begin{eqnarray}
&\Psi_0(r)\to \ln\left[\frac{r}{a_0}\right]\,\textrm{and}&\\
&\Psi_m(r)\to r^m \left[1-\left(\frac{a_m}{r}\right)^{2m}\right],&
\end{eqnarray}
for $m\geq 0$.

The $m=0$ case was studied by Verhaar {\it et al.} \cite{verhaar84} who found the expression
\begin{eqnarray}
\cot\delta_0(k)=\frac{2}{\pi}\left(\gamma+\ln\left[\frac{ka_0}{2}\right]\right)+\frac{1}{2\pi}r_{e}^{2}k^2
\end{eqnarray}
to second order in $k$. We expect a similar expression for the $m=1$ phase-shift, but very little 
can be found on this in the literature. To check this, we compute the exact expression for 
the hard-sphere and for the attractive square well potential. The hard-sphere yields
\begin{eqnarray}\label{hardsphere}
\cot\delta_1(k)=-\frac{4}{\pi(kr_0)^2}-\frac{3}{2\pi}+\frac{2}{\pi}\left(\gamma+\ln\left[\frac{kr_0}{2}\right]\right),
\end{eqnarray}
and the square well gives
\begin{eqnarray}
\cot\delta_1(k)=-\frac{12}{\pi(kr_0)^2}-\frac{11}{2\pi}+\frac{2}{\pi}\left(\gamma+\ln\left[\frac{kr_0}{2}\right]\right),
\end{eqnarray}
in the limit $2\mu r_{0}^{2}V_0/\hbar^2\to 0$ where the depth is $-V_0$. By using the definitions of the 
scattering length, $a_m$, above, we find $a_{1}^2=r_{0}^{2}$ for the hard-sphere and $a_{1}^{2}=\frac{r_{0}^{2}}{3}$
for the square well. The choice of sign for $a_1$ is not given directly by these relations. 
In the regime where there is a bound states in the potential, 
the definition above implies $a_m>0$. Since we know that the hard-sphere potential should not hold a bound
state, a suitable choice is $a_1=-r_0$. Similarly for the square well in the limit  $2\mu r_{0}^{2}V_0/\hbar^2\to 0$
where the centrifugal barrier hinders the formation of a bound state \cite{artem2011a}. We will be 
interested in the regime $a_1>0$ only, as it is nicely comparable to the $m=0$ case. 
To approach the zero-range limit with a bound state always present for the square well, one needs
to take the limit of $r_0\to 0$ and $-V_0\to \infty$ in a manner that keeps $2\mu r_{0}^{2}V_0/\hbar^2$
at or above the critical limit for the appearance of a bound state.

From the discussion above, we get the suggestsive expression for the phase-shift
\begin{eqnarray}
\cot\delta_1(k)=\frac{A}{(ka_1)^2}+B+\frac{2}{\pi}\left(\gamma+\ln\left[\frac{ka_1}{2}\right]\right),
\end{eqnarray}
where $A$ and $B$ are potential-dependent low-energy constants.
We find that the structure of $\cot\delta_1(k)$ is very similar to the $m=0$ case, except that there
is an added $k^{-2}\propto E^{-1}$ term. The leading divergence in the corresponding scattering amplitude 
is therefore a pole rather than the logarithm as for $m=0$. This was pointed out in Ref.~\cite{randeria1990}.
In the main text, we
have shown that in the presence of a harmonic trap, it does not make much difference whether the 
leading $E^{-1}$ term is included or not when calculating the two-body spectrum in the presence
of a bound state. This can be seen from comparison of Fig.~\ref{fig5} (with $A=0$) and Fig.~\ref{fig6}
(with $A<0$). However, in the case with $A>0$ there is no bound state as discussed in the main
text.

\end{document}